\documentclass[12pt]{article}
\usepackage{epsfig,psfrag}

\newcommand{\be}{\begin{equation}}
\newcommand{\ee}{\end{equation}}
\newcommand{\bea}{\begin{eqnarray}}
\newcommand{\eea}{\end{eqnarray}}

\def\id{\protect{{1 \kern-.28em {\rm l}}}}

\def\cN{{\cal N}}

\font\mybbb=msbm10 at 8pt
\font\mybb=msbm10 at 12pt

\def\bbb#1{\hbox{\mybbb#1}}
\def\bb#1{\hbox{\mybb#1}}

\def\C{\bb{C}}
\def\P{\bb{P}}
\def\PP{\bbb{P}}
\def\id{\protect{{1 \kern-.28em {\rm l}}}}

\def\bP{{\bf P^1}}
\def\CO{{\cal O}}
\def\bP{{{\P}^1}}
\def\bC{{{\C}^3}}

\def\id{\protect{{1 \kern-.28em {\rm l}}}}

\makeatletter
\renewcommand\section{\@startsection {section}{1}{\z@}%
                                   {-3.5ex \@plus -1ex \@minus -.2ex}%
                                   {2.3ex \@plus.2ex}%
                                   {\normalfont\large\bfseries}}
\renewcommand\subsection{\@startsection{subsection}{2}{\z@}%
                                   {-3.25ex\@plus -1ex \@minus -.2ex}%
                                   {1.5ex \@plus .2ex}%
                                   {\normalfont\normalsize\bfseries}}
\makeatother

\newcommand{\Tr}{\mathop{{\rm Tr}}}

\begin{document}

\begin{titlepage}

\hfill\hbox to 3cm {\parbox{4cm}{
UCB-PTH-03/03 \\
NSF-KITP-03-10\\
hep-th/0301217
}\hss}

\vspace*{5mm}
\begin{center}


\mbox{\hskip-.8em\Large\bf Massless Flavor in Geometry and Matrix
Models\hss}

\vspace*{8mm}

{Radu Roiban$^a$,~ Radu Tatar$^b$ and~Johannes Walcher$^c$}
\vspace*{5mm}
          

{\it ${}^a$Department of Physics}\\
{\it University of California, Santa Barbara, CA 93106}

\vspace*{3mm}

{\it $^b$Department of Physics, 366 Le Conte Hall}\\
{\it University of California, Berkeley, CA 94720}\\

\vspace*{1mm}

{\it and}\\  
          
\vspace*{1mm}

{\it Lawrence Berkeley National Laboratory}\\
{\it Berkeley, CA 94720}\\

\vspace*{3mm}

{\it ${}^c$Kavli Institute for Theoretical Physics}\\
{\it University of California, Santa Barbara, CA 93106}

\vspace*{3mm}


\end{center}


\begin{abstract}

The proper inclusion of flavor in the Dijkgraaf-Vafa proposal for the solution of 
${\cal N}=1$ gauge theories through matrix models has been subject of debate in the 
recent literature. We here reexamine this issue by geometrically engineering fundamental 
matter with type IIB branes wrapped on non-compact cycles in the resolved geometry, and 
following them through the geometric transition. Our approach treats massive and massless 
flavor fields on equal footing, including the mesons. We also study the geometric transitions and 
superpotentials for finite mass of the adjoint field. All superpotentials we compute reproduce 
the field theory results. Crucial insights come from T-dual brane constructions in type IIA.

\end{abstract}



\vfill
\noindent January 2003
\vskip 1em
\hrule width10em
\vskip .3em
\noindent\hskip .5em
{\small E-mail: 
\parbox[t]{8cm}{\tt ${}^a$radu@vulcan2.physics.ucsb.edu \\
${}^b$rtatar@socrates.berkeley.edu \\
${}^c$walcher@kitp.ucsb.edu}}

\end{titlepage}

\tableofcontents

\section{Introduction}
In the last years many steps have been taken toward a better understanding 
of the dualities between field theory and string theory. One direction was
initiated in \cite{gv} and consisted in studying the large N dualities 
in the context of type A topological strings. 
This topological duality was embedded in the physical superstring theory in 
\cite{vafa} and then further developed  in \cite{CIVA,eot,ckv,civ1,fiol,ook1,cv} (see also
\cite{dot1,dot2,dot3} for an alternative approach, involving brane configurations). 
The main result of these studies was a method of calculating the effective superpotential 
of a four-dimensional field theory using aspects of the flux configurations and of the geometry 
of the compact/transverse space.

It is natural to suspect that similar dualities exist for the type B topological strings 
on Calabi-Yau manifolds. They have been discussed in a series of papers by Dijkgraaf and 
Vafa \cite{dv1,dv2}. On the closed string side of the duality the effective superpotential of the
four-dimensional gauge theory is generated by the Gukov-Vafa-Witten (GVW) 
superpotential \cite{CIVA}. On the 
open string side of  the duality the four-dimensional gauge theory is realized by wrapping 
D-branes on certain 2-cycles and the effective superpotential is generated by the topological 
open string theory living on these cycles. This is described by the holomorphic Chern-Simons 
theory which becomes a simple matrix model with the potential given by the superpotential 
of the gauge theory.
Building on these results, a stronger claim has been argued in \cite{dv3}, stating that for 
a class of 
${\cal N} = 1$ theories, with fields in the adjoint and bifundamental representations of the 
gauge group, the effective superpotential can be expressed in terms of the planar free energy 
of this matrix model.
Using field theory techniques it was shown that, for models with one chiral field $\Phi$ in the 
adjoint representation and a tree level superpotential $W(\Phi)$, the truncation to planar 
diagrams appeared due to the holomorphy of the expected effective action \cite{witten} and/or 
to the cancellation of dependence on momenta between the bosonic and fermionic 
integrals \cite{grisaru}.

It is interesting to extend the original arguments of Dijkgraaf and Vafa to include fields in 
other representations of the gauge group. The case of fields in the fundamental representation
was discussed extensively in \cite{Berenstein}-\cite{feng} (see also \cite{mt} for progress in
other directions), in general without reference to any possible string 
theory realization of such theories\footnote{ Exceptions are references 
\cite{ookuochi} and \cite{hofman}, where the approach considered is
different from the one we use in our paper}. 
It turns out that the matrix/gauge theory relation implies 
that the flavor contribution to the effective superpotential is exactly taken into account by the 
one-boundary free energy of the corresponding matrix model \cite{goteborg,rr,brr,seib}.

It is however rather difficult to extend these results to theories with massless flavor fields. 
Indeed,
in this case the low energy theory is not described only in terms of the glueball superfield as the
naive matrix model predicts, but it must also contain quark bilinears.  An attempt to handle this 
case was introduced in \cite{janik} and requires the introduction of delta functions relating the 
matrix model flavor fields with the corresponding gauge theory mesons. An alternate suggestion 
was presented in \cite{brr} and involves deforming the naive matrix model by mass terms for all 
massless fields and then computing the superpotential from the one-boundary free energy. 
The gauge theory superpotential is obtained by integrating in, in the gauge theory sense, the  
fields that were originally  massless and then taking the massless limit. Finally, these two 
procedures were shown to be equivalent in \cite{feng}. There the gauge theory meson field 
is identified with the Lagrange multiplier enforcing the massless limit.

In this paper we reconsider the geometric arguments which led to the matrix model/gauge theory
duality. First we generalize the results of \cite{CIVA} to the case of an adjoint field of arbitrary 
mass as well as to the case of massless quarks.  As the results of \cite{dv1,dv2,dv3} were based 
on geometric transitions relating open and closed string theories, our new results 
shed a new light on the matrix/gauge theory relation in the presence of massless fields.
We use the T-dual brane configurations \cite{dot1,dot2,dot3,dot4} to understand the dynamics of the
field theory, as the mass of the adjoint field as well as the mass and vevs of the 
fields transforming
in the fundamental representation of the gauge group can be read from the positions of the 
different branes. 

It is important to stress that in our treatment the flavor fields are described by 
D5 branes wrapping  
noncompact two-cycles and these branes exist on both sides of the transition. This is very similar 
to the situations encountered in the analysis of defect CFT-s \cite{DWFO}. On the open string theory 
side the gauge theory is realized on the common part of the world volume of D$5$ branes wrapping
the compact $\P^1$ cycle of the small resolution of the conifold and of D$5$ branes wrapping
the noncompact $\P^1$ cycles. Because of the noncompactness of the  D$5$ branes wrapping
the noncompact  cycles the open strings stretching between them yield no dynamical fields.
On this side of the duality the gauge theory effective superpotential will be generated by the 
dynamics of open strings governed by the holomorphic Chern-Simons theory \cite{AGVA}.

After the geometric transition (which corresponds to the strong coupling limit of gauge theory) 
the D$5$ branes wrapping the compact cycles are replaced by flux.
The branes wrapping noncompact cycles survive the transition and can be interpreted as probes
of the deformed geometry with flux. They give rise to dynamical fields which, roughly 
speaking (we will make this precise later), can be identified with the gauge theory mesons. 
In this formulation of the gauge theory the effective superpotential receives two conceptually 
different contributions. The first part is the flux-generated superpotential while the second 
part is given by the dynamics of open strings starting and ending on the remaining D$5$ branes. 
This latter part reproduces the results of \cite{goteborg} for massive flavor fields as well as the  
ones previously obtained in \cite{OOHO} for the massless ones. 

Finally, we explain the appearance of the delta function identifying the gauge theory meson 
and the matrix model quark bilinear both from a topological string perspective as well as
using the M5 brane dynamics.     

Before proceeding in the following sections with the geometric analysis let us   first summarize 
the gauge theory results we will recover, namely the superpotential for an ${\cal N}=1$ gauge theory 
with both massive and massless quarks, a massive field in the adjoint representation of the gauge 
group and Yukawa interactions.


\subsection{Field theory results for arbitrary mass for the adjoint field \label{fieldth}}

Consider an ${\cal N} = 2$ theory with gauge group $U(N_c)$ and massive and massless quarks 
and consider 
breaking supersymmetry to  ${\cal N} = 1$ by turning on arbitrary  mass term for the adjoint 
chiral multiplet as well as allowing the Yukawa coupling to be different from the gauge coupling.
Denoting the massive quarks by $Q^{(1)}$ and ${\tilde Q}^{(1)}$, the tree level superpotential is: 
\bea
W = \sqrt{2} g \Tr[\tilde{Q} \phi Q] + m\Tr[Q^{(1)}{\tilde Q}^{(1)}]+\mu \Tr \phi^2
\eea
To discuss the Higgs branch of this theory one first integrates out the adjoint field $\phi$. The 
renormalization group fixes the dynamical scale of the resulting theory to be
\bea
\Lambda_{{\cal N} = 1}^{3 N_c - N_f} = \mu^{N_c} \Lambda_{{\cal N} = 2}^{2 N_c - N_f}
\eea
while standard nonrenormalization theorem arguments imply that perturbatively the 
superpotential is just
\bea
\label{massadj}
W = \frac{g^2}{2 \mu} 
\Tr[(Q \tilde{Q} )(Q \tilde{Q})] + m\Tr[Q^{(1)}{\tilde Q}^{(1)}]
\eea
For $\mu \rightarrow \infty$, this superpotential approaches zero and we then obtain
${\cal N} = 1$ SQCD. 

Using symmetry and holomorphy arguments as well as smoothness in the limit 
$g\rightarrow 0$ one can show that, if $N_f \le N_c - 1$, the effective superpotential of this theory 
is just
\be
W = \frac{g^2}{2 \mu} 
\Tr[(Q \tilde{Q} )(Q \tilde{Q})] + m\Tr[Q^{(1)}{\tilde Q}^{(1)}]
+
(N_c-N_f)\left[{\Lambda_{{\cal N} = 1}^{3 N_c - N_f}\over \det(Q \tilde{Q}) }\right]^{1\over N_c-N_f}
\label{fullWeff}
\ee
where the last term represents the nonperturbative contributions. At finite values for the mass 
of the adjoint field and generic values of the mass of the quarks, the expectation value of the 
meson field $M_{ij}=Q_i{\tilde Q}_j$  has  two types 
of diagonal entries \cite{OOHO}. As the mass of the adjoint field is taken to infinity all these 
vacua run away to infinity and there is no vacuum left at finite distance in the moduli space
of ${\cal N}=1$ SQCD with massless quarks.


As stated previously, we will recover the superpotential (\ref{fullWeff}) (and thus all its 
consequences) from geometric considerations. We also find a geometric interpretation of
the `integrating in/out method'' of \cite{INTR}. To achieve this, we will begin by discussing the 
small resolution of the conifold in the case of finite adjoint mass.

\section{Engineering of massive adjoint fields and massless flavors}


In this section we describe the details of the geometric engineering of field theories 
with an adjoint field of finite mass and massive and massless flavor fields. 
We begin by reviewing some results of 
\cite{dot3} concerning the construction of ${\cal N} = 2$ field theories as well as the
addition of fields in the fundamental representation of the gauge group. 
We then proceed to break supersymmetry  by a finite mass for the adjoint field
as well as to find the geometric interpretation of the gauge theory meson field.
In the next section we will discuss the geometric transition of this setup.

\subsection{${\cal N} = 2$ theories from Geometry}

The field theory of interest is realized on the world volume of (fractional) branes
whose transverse space is the tensor product of an ADE singularity with a 
two-dimensional plane. The resolved space contains a collection of $\bP$ cycles, 
together with their normal bundles. For each cycle this is  $\CO(-2) \oplus \CO(0)$. 
The  $\CO(0)$ fibers represent the Coulomb branch of the  gauge theory. Inclusion 
of fields in the fundamental representation of the gauge group as well as breaking 
of supersymmetry to $\cN=1$ by a finite mass parameter for the chiral multiplet 
in the $\cN=2$ vector multiplet is, to some extent, clearer in the brane realization 
of the theory. We will summarize this description which is related to the geometric 
one by $T$-duality. For this we need to examine in slightly more detail the 
geometric description.

The total space of the normal bundle over the $i$-th $\bP$ can be covered with two
patches $\C_{i,S}^3$ and $\C_{i,N}^3$, where $N$ and $S$ refer to the North and 
South pole of the corresponding $\bP$ cycle. The transition functions are given by
\be 
\label{o2}
Z_i'={1\over Z_i}~~~~~~~~Y'_i =Y_iZ_i^2~~~~~~~~X'_i=X_i~~.
\ee
Clearly, the coordinate $X_i$ parameterizes the trivial fibers $\CO(0)$ while 
the remaining coordinates describe the total space of $\CO(-2)\Big|_{\PP^1}$.
To plumb the set of  $\bP$ cycles (together with their normal bundles) and 
reconstruct the full space one uses
certain identifications dictated by the ADE Dynkin diagram associated to the 
chosen singularity.

An $\cN=2$ field theory is constructed by wrapping D5 branes on the $\P^1$ cycles.
The precise interactions of this theory depend on the singularity we started with,
i.e. they depend on the intersection of the $\bP$ cycles. For example, in the 
case of a resolved ${A}_n$ singularity, the $n$ copies of $\CO(-2)\Big|_{\PP^1}$
are connected by the identification
\bea
\label{int}
Y'_i \to Z_{i+1}~~~~~ Z'_i \to Y_{i+1}~~~~~X'_i\to X_{i+1}
\eea
which means that the north pole of the i-th $\bP$ cycle meets the south pole of the
i+1-th $\bP$ cycle. By considering such a singularity together
with  $N_{i}$ D5 branes on the i-th $\bP$ cycle we have a gauge theory  with gauge group 
 $\prod_{i=1}^{n} U(N_i)$ and hypermultiplets $F_i,~\tilde{F_i},i=1,\cdots,n-1$ in the 
bifundamental representations $(N_i, \bar{N}_{i+1})$ and $(\bar{N}_i, N_{i+1})$
respectively, as well as a superpotential consisting of Yukawa interactions of
the fields in $(N_i, \bar{N}_{i+1})$,  $(\bar{N}_i, N_{i+1})$ and 
$(N_{i+1}, \bar{N}_{i+1})$.

To translate the  geometrical picture into a brane configuration 
we split the angular and radial directions of the $\bP$ cycles, and we reduce the
geometrical picture to one where the angular direction is removed. This means 
considering a ``skeleton'' of the geometrical picture. This can be achieved by a 
T-duality \cite{dot1,dot2,dot3}. The T-duality direction is a circle action on the 
normal bundle over the  $\bP$ cycle, given by
\bea
\label{orbitn2}
Z_i \rightarrow e^{i\theta}Z_i,
~~Y_i \rightarrow e^{-i\theta}Y_i\\ \nonumber 
Z'_i \rightarrow e^{-i\theta}Z'_i, ~~Y'_i \rightarrow e^{i\theta} Y'_i~~,
\eea
whose orbits degenerate\footnote{By abuse of terminology we will call these 
degenerate orbits ``lines of singularity''.} 
along $Z_i=Y_i=0$ and $Z'_i=Y'_i=0$. By using \cite{oova}, 
the lines of singularity get mapped into $n$ parallel $NS5$ branes. The $N_i$ D$5$ branes 
wrapped on the blown-up $\P^1_i$ cycle are mapped into $N_i$ D$4$ branes suspended
between the $i$-the and $(i+1)$-th $NS5$ brane.

In the present discussion we are interested in the inclusion of fields transforming in 
the fundamental representation of the gauge group. In the brane realization of the 
theory such fields are introduced by including semi-infinite  D4 branes which 
end on the NS branes. Let us consider that we have $N_{f,i}$ flavors in the fundamental 
representation of $U(N_i)$ and denote these fields by $Q_i, \tilde{Q}_i, i=1,\cdots,n$. 
If the fundamental flavors are massive, we denote their masses by $m_{i}$. They are
given by the distance along the $NS5$ branes between the endpoint of the corresponding 
semi-infinite  D4 brane and the D4 branes describing the $U(N_i)$ part of the gauge 
group.

With this starting point it is easy to construct the geometric version of the setup by 
performing the inverse of (\ref{orbitn2}). The result is that in the geometric picture 
the fields transforming in the fundamental representation of the gauge group
are introduced as D5 branes wrapping non-compact holomorphic 2-cycles given by:
\bea
Y_{i} = 0~~~ \mbox{or}~~~ Y'_{i} = 0,~~~X = m_i  
\eea
The choices $Y_{i} = 0$ or $Y'_{i} = 0$ are identical, as they describe the same 
point in the total space.

\subsection{${\cal N} = 1$ theories from geometry; massless quarks}

We now deform the geometry by adding superpotentials for the adjoint fields (including 
mass terms). Generic theories without matter fields have been analyzed in detail in 
\cite{dot3}. We discuss here the simplest model, obtained by adding just the mass 
term for the adjoint chiral multiplet in the $\cN=2$ vector multiplets, which breaks 
supersymmetry to ${\cal N} = 1$. \footnote{In this section we consider the field theory and 
geometry deformations when the mass for the adjoint chiral multiplet is finite or infinite.
A similar discussion appeared in \cite{gns} for the case of branes probing singularities,
whereas in our case we deal with D5 branes wrapped on blown-up cycles.}
The superpotential will therefore be:
\bea 
\label{supotg} 
W &=&\sum_{i=1}^{n} (\frac{m}{2} \Tr \Phi_i^2\\
&+&\Tr (F_{i} \Phi_{i+1} \tilde{F}_{i}
- \tilde{F}_{i+1}\Phi_{i} F_{i+1} + \lambda_i Q_{i} \Phi_{i} \tilde{Q}_{i} +
\lambda'_i Q_{i+1} \Phi_{i+1} \tilde{Q}_{i+1}))
\nonumber
\eea
where $\lambda_i$ and $\lambda'_i$ are $\sqrt{2}g$ if the Yukawa interactions are to 
preserve ${\cal N} = 2$ supersymmetry, but can have arbitrary values for the 
${\cal N} = 1$ theories. 

For a better understanding we begin by considering a theory with gauge group
$U(N)$ and $N_f$ fields in the fundamental representation, with mass 
parameters $m_i$. Before supersymmetry breaking, this is the world volume 
theory of $N$ D5 branes wrapped on the nontrivial
$\P^1$ cycle of a blown up $A_1$ singularity and $N_f$ D5 branes wrapped on 
the noncompact cycles defined by
\bea
Y = 0,~X=m_i~~~~i=1,\dots,m
\eea
or 
\bea
Y' = 0,~X=m_i~~~~i=1,\dots,p
\eea
with  
\be
m+p=N_f~~.
\ee
The brane configuration corresponding to this geometry is constructed out of two parallel
$NS5$ branes with $N$ D4 branes suspended between them as well as $m$ and $p$
semi-infinite D4 branes ending on the left and right $NS5$ brane, respectively. In this 
language supersymmetry breaking is realizes by rotating the $NS5$ branes relative to 
each-other. The rotation angle is a function of the mass of the adjoint field.

The T-duality described in the previous section provides the connection between 
the rotated brane configuration and geometry. Roughly speaking, rotating the 
$NS5$ branes corresponds to fibering the $A_1$ singularity over the dimensional plane.
In other words, the normal bundle of the blown up $\P^1$ is modified. The fields
transforming in the fundamental representation are still described by D5 branes 
wrapping noncompact cycles. Unlike the situation above, after the rotation the two 
choices  of cycles become physically inequivalent.

Two limits of geometry as a function of the mass of the adjoint are important to discuss:

$\bullet$ $\bP$ with normal bundle $\CO(-2) \oplus \CO(0)$, obtained for a 
zero mass for the adjoint field. In this limit the field theory has ${\cal N} = 2$
supersymmetry, as discussed in the previous section.

$\bullet$ $\bP$ with normal bundle $\CO(-1) \oplus \CO(-1)$ (the resolved conifold), 
obtained for an infinite mass for the adjoint field. In this limit the field theory in the
world volume of the D$5$ branes is ${\cal N} = 1$ SQCD limit, i.e. the field in the adjoint
representation is decoupled.

This latter choice is described by two  copies of $\bC$, parametrized by $(X,Y,Z)$ and $(X',Y',Z')$,
together with the transition function:
\bea
Z' = Z^{-1},~X' = X Z,~Y'=Y Z~~.
\eea
As discussed in \cite{dot1}, the singular conifold is recovered through the blowdown map
\bea
x = X = X'Z',~y=Z Y~=Y',~u=~Z X=~X',~v=Y=Z' X'
\label{blowdown}
\eea
which implies that
\be
x y - u v = 0
\label{singconif}
\ee
which defines the conifold at the singular point. This map together with (\ref{orbitn2}) 
induce a circle action on the coordinates in the two patches which can be used to 
translate between the brane and geometric description:
\bea
\label{orbitn1}
Z \rightarrow e^{i\theta}Z&,& ~X \rightarrow X~~,~~~~
~~Y \rightarrow e^{-i\theta}~Y\\ \nonumber 
Z' \rightarrow e^{-i\theta}Z'~~, &{}& X' \rightarrow e^{i\theta}~X'~~,~~~~Y' \rightarrow Y'~~.
\eea
The lines of singularity are $Z = Y = 0$ in the first $\bC$ and $Z' = X' = 0$ in the second
$\bC$, which are clearly orthogonal. Thus, the brane configuration corresponding to 
the small resolution of the conifold contains two orthogonal $NS5$ branes.
 
Let us now analyze the fields in the fundamental representation. As we discussed before,
they correspond to D5 branes wrapping non-compact holomorphic cycles  
\bea 
Y = 0,~~X =m 
\eea
or to D5 branes wrapping a non-compact holomorphic 2-cycles 
\bea 
X' = 0,~~Y' = m~~.
\eea
Thus, we notice that after the $A_1$ singularity was fibered over the transverse 
two dimensional space, these two cycles are no longer equivalent. 
Indeed, as the lines of singularity in the geometry are now along the $X$ and $Y'$ directions, 
after a T-duality on the above orbit we get 
two orthogonal NS branes on the directions $X$ and $Y'$. The D5 branes wrapped 
on the compact $\bP$ are mapped into finite D4 branes (between the two 
orthogonal NS branes) while the ones on the non-compact holomorphic cycles  
map into semi-infinite D4 branes which can end on one NS brane or the other. 

There are also ${\cal N} = 1$ brane configurations (and geometries) which correspond to 
finite masses for the adjoint field. As we stated above, in terms of brane configurations
this means that the NS branes are neither parallel nor orthogonal. By a T-duality  we 
can add a circle to this ``geometric skeleton'' and obtain a geometry where the lines of 
singularity are neither parallel nor orthogonal. To make this more concrete, the 
transition function $X' = X Z$
is replaced with $X' = X_r Z$ where $X_r$ is some  function of $X,~Y$ and $Z$.
Thus, the geometry is now:
\bea
Z' = \frac{1}{Z},~~X' = X_r Z,~~Y' = Y Z
\eea 
Taking
\be
X_r=X-{1\over m_{\it adj}} YZ
\ee
and using the blowdown map (\ref{blowdown}) we find the following deformation 
of the singular conifold:
\be
uv-y(x-{1\over m_{\it adj}}y)=0~~.
\label{modeq}
\ee
In the limit of infinite $m_{\it adj}$ we recover the usual conifold geometry while
in the limit of vanishing $m_{\it adj}$ rescaling $u$ and $v$ leads to the $A_1\times \C$.

The orbit (\ref{orbitn1}) has the form:
\bea
\label{orbitrot}
Z \rightarrow e^{i\theta}Z~, ~~X_r \rightarrow X_r~,
~~Y \rightarrow e^{-i\theta}~Y \\ \nonumber 
Z' \rightarrow e^{-i\theta}Z'~, ~~X' \rightarrow e^{i\theta}X'~,~~Y' \rightarrow Y'~~,
\eea
and we observe that the degeneration is indeed along the union of complex lines 
along $X_r$  in the first $\C^3$ and $Y'$ in the second $\C^3$. As promised, T-duality
on this orbit produces a configuration of two $NS5$ branes at an angle determined by $X_r$.

We can reach a similar result by starting with the ${\cal N} = 2$ geometry
(\ref{o2}) (with $i=1$) and deforming the transition functions to
\bea
\label{deform}
Z'= 1/Z~, ~~~~Y' = YZ^2 + m_{A} X Z~~.
\eea 
To see what happens when we vary the mass of the adjoint field, we switch again to 
the map (\ref{blowdown}) and find
\bea
\label{fadj1}
u v - y^2 + m_{A} x y = 0~~,
\eea
which is the same equation we had before, up to a rescaling of $u$ and $v$.

The geometric transition takes us to a deformed conifold. Since there exists a 
holomorphic change of coordinates which casts equation (\ref{fadj1}) into that of the
conifold, one may say that the two geometries describe the same physics. This is, 
however, not the case as various boundary conditions change under these 
transformations. Anticipating later arguments, the 
boundary conditions can be naturally chosen in one coordinate system while the 
computations are easier in the other one; the coordinate 
transformation will introduce a dependence on the mass of the adjoint field in the
boundary conditions.

Up to now we have discussed the geometric construction of massive fields in the 
fundamental representation of the gauge group. Our main goal is, however, to find a 
geometric description of massless matter fields. To reach this goal we start with
the brane configuration describing such fields \cite{hw} and then subject it to
T-duality transformations along the orbit (\ref{orbitrot}). 

There are two choices of introducing matter, one with semi-infinite D4 branes and 
the other with D6 branes. In the following we will use D$4$ branes for this purpose
and begin by describing the setup at vanishing string coupling, when all 
branes are represented by straight hyperplanes. 

\begin{figure}[ht]
\begin{center}
\psfrag{a}{D$4_m$}
\psfrag{b}{D$4_M$}
\psfrag{c}{D$4_c$}
\psfrag{m}{$m$}
\psfrag{M}{$M$}
\psfrag{al}{$\alpha$}
\psfrag{e}{$89$}
\psfrag{f}{$6$}
\psfrag{g}{$45$}
\psfrag{k}{${\it NS5}$}
\psfrag{l}{${\it NS5}'$}
\epsfig{file=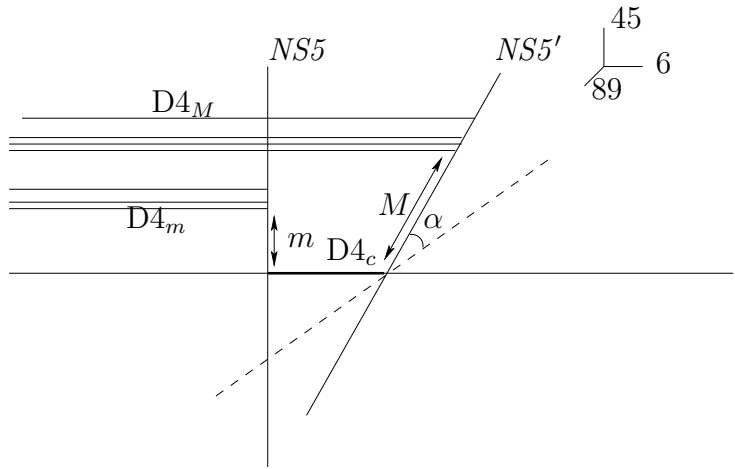,height=6cm}
\caption{ Brane construction .\label{branes}}
\end{center}
\end{figure}

To be specific, we consider the configuration in figure \ref{branes}.
The dashed line represents the directions orthogonal on the $(456)$ space. The fact that 
the $NS'$ brane is not orthogonal on this 3-space corresponds in field theory language 
to introducing a superpotential quadratic in the chiral superfield transforming in the 
adjoint representation of the gauge group. This can be easily seen using symmetry 
arguments \cite{OOHO,BARB}. The $N=2$ theory is invariant under 
$U(1)\times SU(2)$ $R$-symmetry
which correspond to rotations in the $(45)$ and $(789)$ directions, respectively. In the rotated 
brane configuration $SU(2)$ is broken to $U(1)$ corresponding to rotations along the $NS'$ directions.
Denoting by $x$ the coordinate along the $NS$ brane and by $y$ the coordinate along the $NS'$ one,
it follows that they are proportional and that the proportionality coefficient is charged under both
$U(1)$ $R$-symmetry groups, with charges of the same magnitude and opposite sign. The
only field theory object with these properties is the mass of the adjoint field.

Let us now discuss the interpretation of the position of the end of the D$4$ branes on the $NS$ 
and $NS'$ branes. Separating the D$4$ branes induces a breaking of the $U(N_f)$ flavor symmetry.
If the separation is in the $(4,5)$ direction, then it should be interpreted as a field theoretic mass term
for the quarks as this is the only parameter charged under the corresponding $U(1)$ symmetry.
If the separation is along the $NS'$ brane this should be interpreted as breaking due to a 
nonvanishing eigenvalue for the meson field. Indeed, this is the only field theoretic object charged 
under the second $U(1)_R$ symmetry.

At vanishing string coupling, it is possible to understand
from figure \ref{branes} the phenomena which occur when one changes the 
description of the theory from having a massive field to having a bilinear in that field
with  non vanishing expectation value. Indeed, all one has to do is to transform a D$4_m$ 
brane into a D$4_M$ one. This is possible only by recombining the D$4_m$ brane with one 
of the color branes. Under this operation the gauge group is spontaneously broken, as
at vanishing (string) coupling the only way for a bilinear in fields to have vev is for each of the 
two factors to have a vev. Thus, in the process of changing the description of the theory 
from having a massive field to having a bilinear with non vanishing vev, the rank of the 
unbroken gauge group decreases with the number of  D$4_m$ branes transformed into D$4_M$. 
This recombination of different types of D4 branes can be interpreted as the analog of the 
field-theoretic ``integrating in/out'' procedure of \cite{INTR}. 

At finite coupling it is certainly possible for a composite operator to acquire a non vanishing vev
without its building blocks having one. However, if the vev is larger than the dynamical scale of the 
theory, the vev can be treated classically and thus, in the case of quark bilinears, leads to 
a spontaneous breaking of the gauge group as well. We will return to this in a later section
and quantitatively recover this picture from a geometric description.

This configuration can be easily mapped  to the type IIB configuration. As discussed earlier,
the $NS$ branes and the compact D$4$ branes are mapped to the resolved conifold in 
coordinates (\ref{modeq}); there are two lines of  singularity emerging from the north and 
south pole of $\P^1$, the angle between them being given by the mass of the adjoint field.
A set of non-compact $\P^1$ cycles with D$5$ branes wrapping them end on these lines. 
Depending on their orientation,
their end point on the north pole line describes the mass of the corresponding field
while the end point on the south pole line describes a vev, or vice versa. 

As any geometry containing a conifold singularity, this one exhibits a geometric transition 
similar to the standard one. Because of the presence of the D-branes on the 
non-compact cycles there is  more information which needs to be taken through
the transition.

\section{Geometric transition with fundamental fields \label{funnyS}}

The issue of introducing fundamental matter in the geometric transition has been considered in
\cite{CIVA,dot2,ookuochi}; however, the analysis applies only to a theory with all quarks integrated 
out (i.e. they are all massive) and only for an infinitely massive ${\cal N} = 2$ adjoint field. 
The goal of the present section is to relax these assumptions and recover the much richer set of
results described in section \ref{fieldth}. In particular, since the low energy theory is described 
in terms of mesons, an essential ingredient is their geometric interpretation.

There are two different ways of describing this. One way is to start from the 
brane configuration described above, lift it to $M$-theory and then map the results to the 
deformed geometry, paying particular attention to the vev-s of the fundamental fields. 
An alternative way is to start from a brane configuration describing a field theory
with bifundamental fields which reduces to the theory of interest in a certain limit, 
describe the transition for its associated geometry and then take the appropriate limit
at the end. The issue of vev-s for the bifundamental
fields was discussed for the brane configurations in \cite{bh} and in the context of the geometric 
transition in \cite{dot3}. The former line of reasoning  emphasizes the behavior of the flavor branes 
under the geometric transition while somewhat obscuring the precise identification of the
vevs of the fundamental fields. The  latter argument identifies more clearly the vev of the 
fundamental fields in the deformed geometry while somewhat obscuring the behavior of 
flavor branes under the transition.  For these reasons as well as for others which will become 
clear in section \ref{mm}, we will describe both approaches.

Starting with the brane configuration in figure \ref{branes}, the strong coupling limit is understood
as lifting it to $M$-theory together with taking the separation between the two $NS5$ branes to zero.
Arguments similar to those in \cite{dot1}-\cite{dot4} imply  that the brane configuration becomes an M5 
brane with the world volume \cite{wit,OOHO,GIKU} given by the 
curve
\be
y {\hat x}=  \zeta  ~~~~{\hat x}=x-{y\over 4m_{\it adj}}
\label{curveM}
\ee
where ${\hat x}$ and ${y}$ are the coordinates along the two $NS5$ branes while $x$ is the coordinate
along the dashed line in figure \ref{branes}. As $x\rightarrow\infty$ we have either $y=0$ or 
$y\rightarrow m_{\it adj} x$. In these coordinates $y$ is the coordinate along the dashed line, 
$y=x_8+i x_9$, while $x$ is the coordinate along $NS5$, $x=x_4+ix_5$. The coefficient of $y^2$ can be 
freely adjusted to any non vanishing value, while remaining proportional to the mass 
of the adjoint field. We will use this freedom to identify $m_{\it adj}$ 
with the mass parameter of the adjoint field.  

The M theory lift of a D$4$ brane describing a field in the fundamental representation 
is a cigar-shaped $M5$ brane which intersects the curve (\ref{curveM}) in exactly one point, say $P$. 
As this point is constrained to lie on (\ref{curveM}) only one of its coordinates can be arbitrarily chosen. 
This is consistent with the field theory expectation that, given the superpotential and a mass parameter
there is a discrete set of choices for the expectation value of the meson field. Conversely, given the 
superpotential and an expectation value of the meson field, the mass parameter is uniquely 
determined. In type IIA language, this represents a set of semi-infinite D$4$ branes ending on an
$NS5$ brane whose world volume is given by (\ref{curveM}).

The discussion in the previous section suggests that the coordinate of  $P$ along 
the $x$ direction equals the mass of the corresponding field while its coordinate 
along the $y$ direction represents the expectation value of the meson field built out of the 
corresponding field. Therefore, the strong coupling/M-theory analog of the 
``integrating in/out'' procedure of \cite{INTR} represents the transition between the two choices 
of which one of the two coordinates of the point $P$ is fixed as ``boundary condition''. 

We want to emphasize that during the transition only the compact $\bP$ cycle shrinks, but 
the non-compact 2-cycles remain unchanged. In type IIA theory this represents the
fact that at strong coupling there still exist semi-infinite D4 branes 
which end on the $NS5$ whose  world volume is the curve (\ref{curveM}). 

Let us now turn to the other description of the geometric transition with flavor fields.
The starting point is a theory with $U(N_c)\times U(N_f^m)\times U(N_f^M)$ 
as gauge group. An ${\cal N}=1$ brane configuration realizing this theory involves 
three $NS5$ branes, say A, B and C, at different points in the $x^6$ direction, 
whose projection on a $(x,y)$ plane forms a triangle, with corners denoted by
$I_{AB}$, $I_{AC}$ and $I_{BC}$, respectively. Along the $x^6$ direction 
and at each corner of this triangle lie $N_c$,  $N_f^m$ and $N_f^M$ D$4$ 
branes, respectively.  For definiteness, let us assume that there are $N_c$
between B and C,  $N_f^m$ between A and B and $N_f^M$ between A and C.
This brane configuration was analyzed in detail in \cite{GIPE} where it was 
obtained by rotating an ${\cal N}=2$ brane configuration describing a gauge  theory 
with gauge group $U(N_c+N_f^M)\times U(N_f^m+N_f^M)$ and bifundamental fields. 
Among other things, it was shown that the distance measured along the B brane 
between $I_{AB}$ and $I_{BC}$ is equated with the mass of the bifundamental 
fields while the distance {\it measured in the direction orthogonal to the
{\rm B} brane} between $I_{BC}$ and $I_{AC}$ is equated with the vev of the 
off-diagonal components of the scalar field in the adjoint representation which 
break $U(N_c+N_f^M)$ to $U(N_c)$.

To recover the brane configuration described in the previous section we take the 
A brane to infinity in the $x^6$ direction, without crossing the other NS branes (in
what follows, we denote this process as a decoupling limit) ;
in this limit the $U(N_f^m)$ and $U(N_f^M)$ gauge bosons become nondynamical,
the gauge symmetry becomes global. Thus, the bifundamental fields survive
as fundamentals of $U(N_c)$.

This setup was described geometrically in \cite{dot3} in terms of a resolved $A_2$ 
singularity fibered over a plane. Among other things, it was shown that in the slices of fixed 
$x^6$ and $x^7$ there exists a 1-cycle and the inverse image under the projection onto these slices 
of the compact domain bounded by it is homotopic to an $S^3$.
It was also shown that this cycle exists on both sides of the geometric transition.
Since its size is proportional to the expectation value of the bifundamental 
fields, it can be used to give an invariant meaning for this expectation value in the deformed 
geometry\footnote{The results of \cite{dot3} describe the existence of two types of deformations 
in the deformed geometry. The first type are the ``normalizable deformations'' and correspond 
to dynamical quantities in field theory (e.g. the glueball superfield). The second type are the 
``non-normalizable deformations'' and correspond to non-dynamical quantities in field 
theory (as the vev of the bifundamental fields).}.

The geometric version of the fact that the brane A is taken to infinity is that the leftmost 
singularity line is
taken to infinity without crossing the other two lines. In this limit two of the three $\P^1$
cycles decompactify and we recover the geometry described in the previous section. It is also clear 
that this limit can be taken as long as no geometric transition occurs for the 2-cycles which 
decompactify. Indeed, the geometric transition is the geometric version of the strong coupling
limit while the decompactification is the geometric image of a small coupling limit.

In the resolved geometry, the decoupling limit leads to a degeneration of the ``non-normalizable'' 
$S^3$ cycle into an infinitely thin and infinitely long submanifold  which touches all $\P^1$ cycles. 
Its projection onto the left-most line of singularities describes the mass of the massive flavor fields
while its projection onto the direction orthogonal to it is proportional to the meson expectation value.

If the decoupling limit is taken after a geometric transition occurs for the $\P^1$ cycle 
between the two rightmost singularity lines, the ``non-normalizable'' $S^3$ cycle degenerates into 
an infinitely thin and infinitely long submanifold  which touches the  special Lagrangian cycle
and the noncompact $\P^1$ cycles. 

In terms of brane configurations, the two pictures correspond to moving the brane A to 
infinity before or after the $NS5$ branes B and C are deformed into a unique one. From this
point of view it it clear that the decoupling limit and the deformation of the $NS5$ branes commute.
In geometric terms the two pictures correspond to decompactifying two $\P^1$ cycles
before and after a geometric transition occurs for the third one; here, the decompactification 
commutes with the geometric transition because of the geometric nature of each process. 

Comparing the two pictures implies that the mass of the massive fields is given by the
distance between the special Lagrangian cycle and the noncompact $\P^1$ cycle ending above the 
point $I_{AB}$. Denoting this direction by $x$, the vev of the remaining fields transforming in the 
fundamental representation of the gauge group is given by the 
projection of the distance between the special Lagrangian cycle and the noncompact 
$\P^1$ cycle ending above the point $I_{AC}$ onto the normal to $x$. This sharpens the 
identifications suggested by the first description of flavor fields.

\section{Effective superpotential at strong coupling  \label{computations}}

After having discussed all the details of the geometric transition for an adjoint field of finite mass
as well as for massive and massless flavor fields let us proceed to the computation of the effective 
superpotential. In this section we  recover the gauge theory results (\ref{fullWeff}) (or rather 
their form when the glueball superfield is included) from the deformed geometry with branes and fluxes.
In the next section we will find the same results by expressing the computations in terms of a matrix 
model.

Thus, the starting point is a closed string background given by the deformed conifold 
in the coordinates in which its defining equation is:
\be
pq+y (x-{y\over 4m_A})=  \zeta~~.
\label{stwcon}
\ee
In this geometry there exist D$5_m$ branes wrapping the noncompact cycles
defined by the equation $q=0$ and boundary condition\footnote{As described 
in \cite{AGVA}, it is necessary to impose a boundary condition only in one of the $x$ 
or $y$ direction, as the other one is determined by equation (\ref{stwcon}).} 
$x(p\rightarrow\infty)=x_*$ as  well as D$5_M$ defined by the same equation $q=0$ 
but a different boundary condition $y(p\rightarrow\infty)=y_*$.

As discussed in \cite{CIVA},
the superpotential of the gauge theory dual to a 
configuration of fluxes and branes consists of two parts. The first part represents the 
contribution of fluxes and it is given by the GVW superpotential:
\be
W_F=\int\Omega\wedge F~~.
\label{GVW}
\ee
The second part consists of the contribution of branes. The theory living on the part of
the branes wrapping the cycles is given by the holomorphic Chern-Simons action \cite{WIT92}. 
Their contribution to the  superpotential can be computed by evaluating this action on  
a (generic) classical field configuration. This operation, which essentially integrates 
out at the classical level the fluctuations around the classical solution, describes the 
obstructions to the deformation of the branes. As we are interested in evaluating 
this action on a non-compact brane, the boundary conditions at infinity are kept fixed.

In the context of the conifold geometry describing an infinitely massive adjoint field and for 
branes describing massive flavor fields, both these contributions 
were computed in \cite{CIVA}. We will extend this computation to describe 
a finite mass parameter for the adjoint field as well as flavor fields which develop large
expectation values for their corresponding mesons.

To evaluate the  superpotential (\ref{GVW}) one usually writes it in terms of
periods of $\Omega$ as well as fluxes through the dual cycles.  
\be
W_F=\int_A\Omega\int_B H_{NS}-\int_B\Omega\int_A F_{RR}=
\tau S - N_c \Pi
\ee
where 
\be
S=\int_A\Omega~~~~~~~~\Pi=\int_B\Omega~~~~~~~~\tau=\int_B H_{NS}
~~~~~~~~N_c=\int_A F_{RR}
\ee

The periods of $\Omega$ over compact cycles are invariant under changes of 
coordinates which do not change the complex structure. Thus, it is easy to see that 
the relation between $S$ and the deformation parameter $\zeta$ is identical to the one 
in the case of infinite mass for the adjoint field. Indeed, introducing the coordinates
\be
u=  \sqrt{m_A} x ~~~~~~~~v= {y\over 2\sqrt{m_A}} - \sqrt{m_A} x
\label{coord}
\ee
one can write the equation (\ref{stwcon}) as the usual big resolution of the conifold:
\be
u^2-v^2=\zeta
\ee
This change of coordinates is holomorphic.
Therefore, by writing the cycle as a 2-sphere fibered over a segment, we find \cite{CIVA}:
\be
S=\int_A \Omega = \int\limits_{-\zeta^{1/2}}^{\zeta^{1/2}}  \sqrt{u^2-\zeta}\,du = 
{\zeta\over 4}
\ee

This is however not the case for periods over non-compact cycles. Indeed, the 
corresponding integrals are defined with a cutoff which changes under 
coordinate transformations. In the $(u, v)$ coordinates, the $B$-cycle can be defined as 
an $S^2$ fibration over a curve starting\label{the starting point is fixed by the 
fact that the intersection of the $A$ and $B$ cycles is unity.} at $u=\sqrt{\zeta}$ 
and ending at some cutoff.
However, we are interested in the periods computed in the $(x,y)$ coordinates and thus 
the cutoff in the  $u$-plane should be derived from a more fundamental cutoff
in the $x$-plane. The two cutoffs are related by (\ref{coord}); thus, the period integral 
defining $\Pi$ for finite mass for the adjoint field is:
\be
\Pi=\int\limits_{\zeta^{1/2}}^{\Lambda_0 \sqrt{m_A}} \sqrt{u^2-\zeta}\,du =
{1\over 2}\Lambda_0^2m_A +
\left[-{1\over 4}\zeta - {1\over 4}\zeta\ln{{\Lambda_0^2m_A\over S}}
\right]+{\cal O}({1\over \Lambda_0})
\ee

Ignoring the terms which are polynomially divergent as the cutoff is taken to infinity,
it follows that  the GVW superpotential is
\be
W_F = S\left(\ln {\Lambda^{2N_c}m_A^{N_c}\over S^{N_c}}+N_c\right)
\label{puregauge}
\ee
where we also used the usual definition for the dynamical scale in terms of the cutoff and 
the gauge coupling (a.k.a. ``dimensional transmutation'')
\be
\Lambda^{2N_c}=e^{-\tau}\Lambda_0^{2N_c}
\ee

Equation (\ref{puregauge}) is indeed the correct gauge-theoretic expressions
for energy scales less than $m_A$: the adjoint field is integrated out and its mass contributes 
to the dynamical scale:
\be
\Lambda_{\cN=1}^{3N_c}=\Lambda_{\cN=2}^{2N_c}m_A^{N_c}
\ee

We now turn to the contribution of the D$5$ branes describing the fields charged under
global symmetry groups. As described above, they contribute to the effective 
superpotential an amount equal to the holomorphic Chern-Simons action (which 
is the theory living on the part of the brane wrapping the cycle) evaluated
on a representative of the homology class of 
the non-compact  2-cycles with generic moduli dependence.

As in the case of the $B$-cycle described above, a proper definition for these cycles 
requires a choice of boundary  conditions\footnote{More formally, they are cycles 
in a nonstandard relative 
homology group, for which the constraint is given by the boundary conditions.}. 

To begin with, we recall that the Calabi-Yau space of interest is given by
\be
pq=F(x, y)
\ee
embedded in $\C^4$. In this space, the noncompact cycles we are interested in
are defined by \cite{AGVA}:
\be
{\cal C}:~~~~F(x, y)=0~~~q=0~~~~x(p\rightarrow\infty)=x_*~~~
y(p\rightarrow\infty)=y_*
\ee
where $x_*$ and $y_*$ represent boundary conditions and the function $F(x, y)$ is 
given by the right-hand-side of the equation (\ref{curveM}). 
The coordinate parameterizing the cycle is denoted by $p$ while 
the position of the cycle in the total space is described by a point $(x_*,\,y_*)$ on 
the curve $\Sigma:~F(x, y)=0$.

In \cite{AGVA} it was shown that the holomorphic Chern-Simons action can be 
written as:
\be
S=\int_{\cal C} {dp\over p} \alpha\,d\beta = -\int_{\cal C} {dp\over p} \beta\,d\alpha 
\ee
where $\alpha$ and $\beta$ are coordinates parameterizing the curve $\Sigma$ and 
describe one of the directions the cycle is allowed to fluctuate in. This 
in turn implies that only one of them can be chosen independently as boundary condition;
the other one is determined by the requirement that $(\alpha,\,\beta)$ 
lies on $\Sigma$. As the integral over $p$ factorizes, we are left with
\be
S=\int_{\Sigma} \alpha\,d\beta = -\int_{\Sigma}  \beta\,d\alpha 
\label{hCS}
\ee

In choosing the boundary conditions at infinity we have to make sure that they 
represent a stable point on the curve at infinity. In other words, the intersection 
point between the cycle and $\Sigma_*=\Sigma(p\rightarrow\infty)$ should
be one of the critical points on the direction along which boundary conditions are chosen.

With these clarifications, let us now evaluate (\ref{hCS}) for D$5_m$. In $(x,\,y)$ 
coordinates, their position on the $x$ axis near the origin of the coordinate along 
the cycle describes the mass of the corresponding quarks.  Thus, it is natural to 
fix the boundary conditions at infinity in these coordinates. We will nevertheless 
evaluate the action in the $(u,\,v)$  space. 

As explained above, we fix the $x_*$ such that it is a critical point of $y(x)$. Fixing 
the origin on $\Sigma_*$ at $x=m$ and solving $F(x, y)=0$ for $y$ we find that one 
of the critical points is at infinity, which we 
regularize by introducing a cutoff $\Lambda_0$. Translating to the initial origin
on $\Sigma$ we find that we must integrate over
\be
x\in [m,\,\Lambda_0+m]~~.
\ee
This interval can easily be translated into an integration domain for $u$. Ignoring
terms which are polynomially divergent as the regulator is removed as well as terms 
which vanish in this limit, it follows that
the superpotential is:
\bea
W_m&=&{1\over 2}
\int\limits_{m\sqrt{m_A}}^{(\Lambda_0+m)\sqrt{m_A}} \sqrt{u^2-\zeta}\,du 
\label{wm}\\
&&\!\!\!\!\!\!\!\!\!\!\!\!\!\!\!\!\!\!
=
-S\left[{1\over 2}+{1\over 4k_mS}\left(\sqrt{1-{4k_mS}}-1\right)
-\ln({1\over 2}+{1\over 2}\sqrt{1-{4k_mS}})\right]+S\ln{m\over \Lambda_0}
\nonumber
\label{massive}
\eea
where we have introduced the notation
\be
k_m={1\over m^2m_A}~~.
\ee 

We can easily recover the results of \cite{AGVA} by taking the mass of the adjoint 
field to infinity, or rather equal to the cutoff. It is not hard to see that the only 
surviving term from the equation above is
\be
W_m(m_A\rightarrow\infty)=S\ln{m\over \Lambda_0}~~.
\ee

Let us now consider the contribution of D$5_M$to the effective superpotential. As 
in the previous situations, we will fix the boundary conditions in the $(x,\,y)$ space and
then translate them to $(u,\,v)$. In the previous section we argued that
the projection on the $y$ axis of the displacement of the D$4_M$ brane along $NS5'$
can be identified with the meson expectation value. Thus, we will consider a
noncompact 2-cycle which ends at coordinate $y=4\sqrt{2}iM$.~\footnote{The numerical 
factor can be traced to a similar factor in equation (\ref{curveM}). It is related to a 
different choice of $y$ coordinate compared to \cite{OOHO}.}

Determining the boundary 
condition at infinity is slightly more involved. First we solve the  equation
(\ref{curveM}) for $x(y)$:
\be
x={y^2 + 4\zeta^2m_A^2 \over 4m_A y}~~.
\ee
From here we see that there are several critical points. To make connection with 
the brane picture, we would like to pick boundary conditions such that, as the coupling 
constant is decreased, the mesons will have a large expectation value. As there is no 
critical point at infinity for real values of $y$, we will 
choose the brane to end at the critical point at imaginary infinity in the $y$ direction. 
\be
y_*= 4\sqrt{2}i(\Lambda_0^2)~~~~~~x(y_*)=4i{\Lambda_0^2\over 4m_A}
+{\cal O}({1\over \Lambda_0})~~.
\ee
Therefore, we have the following integration domain:
\be
y\in 4\sqrt{2} [i\Lambda_0^2,\,iM]
\ee
which in $(u,\,v)$ coordinates becomes
\be
v\in {\sqrt{2}\over\sqrt{m_A}} [i{\Lambda_0^2},\,i{M}]
\ee
upon assuming that $M$ is large. 

Thus, we are required to compute:
\bea
W_M&=&{1\over 2} 
\int\limits_{i{\sqrt{2}\Lambda_0^2\over \sqrt{m_A}}}^{i{\sqrt{2}M\over \sqrt{m_A}}}
\sqrt{v^2+\zeta}\,dv=
-{1\over 2} 
\int\limits_{{\sqrt{2}\Lambda_0^2\over \sqrt{m_A}}}^{{\sqrt{2}M\over \sqrt{m_A}}}
\sqrt{v^2-\zeta}\,dv\nonumber\\
&=&-{M^2\over 2 m_A}+S\ln{\Lambda_0^2\over M} 
\label{wM}
\eea
where we have ignored terms of order ${1\over \Lambda}$ and $1\over M$.

We are now in position to construct the full superpotential. However, in combining
equations (\ref{puregauge}), (\ref{wm}) and (\ref{wM}) we have to be careful in 
counting the $RR$ flux through the $A$-cycle. As we saw in 
an earlier section, for small
coupling, an expectation value for the meson field is equated with an expectation value 
for the fundamental fields. If the expectation value of the meson is larger than the 
dynamical scale but smaller that the cutoff, a similar identification is 
possible\footnote{Since the quantum contribution to the expectation value of the meson
is equal to the dynamical scale up to coefficients of order one it follows that, if  it is larger than $\Lambda$, 
it must be generated at the classical level}. Since 
we assumed $M$ to be large and comparable to $\Lambda_0^2$, this is the regime 
we are studying. Thus, the brane picture applies without modification and the rank of the
gauge group is smaller compared to the pure gauge theory by an amount equal to the 
rank of the expectation value of the meson matrix. The full superpotential is therefore:
\bea
W_{\it full}\!\!&=&\!\!
S\left(\ln {\Lambda_0^{2(N_c-N_f^M)}m_A^{N_c-N_f^M}\over S^{N_c-N_f^M}}
+N_c-N_f^M\right)-\tau S\nonumber\\
&+&\!\!S\ln {\Lambda_0^{2N_f^M}\over \det M} - {1\over 2 m_A}\Tr[M^2]
+S\ln{\det m \over \Lambda^{N_f^m}_0}\nonumber\\
&-&\sum_m 
S\left[{1\over 2}+{1\over 4k_mS}\left(\sqrt{1-{4k_mS}}-1\right)
-\ln({1\over 2}+{1\over 2}\sqrt{1-{4k_mS}})\right]\nonumber\\
&=&\!\!-{1\over 2 m_A}\Tr[M^2]+
S\left(\ln {\Lambda^{3N_c-N_f}\det m\over S^{N_c-N_f^M}\det M}
+N_c-N_f^M\right)\nonumber\\
&-&\!\!\sum_m 
S\left[{1\over 2}+{1\over 4k_mS}\left(\sqrt{1-{4k_mS}}-1\right)
-\ln({1\over 2}+{1\over 2}\sqrt{1-{4k_mS}})\right]
\label{Wfull}
\eea
where $k_m=m^2m_A$, $\Lambda$ is the dynamical scale at which the adjoint field 
is integrated out, $N_f^M$ is the number of fundamental fields combined into mesons,  
$N_f^m$ is the number of massive fundamental fields and $N_f=N_f^M+N_f^m$.

\section{Effective superpotential at weak coupling; Matrix Models \label{mm}}

In this section we recover the field theoretic effective superpotential in the resolved 
geometry and provide a geometric justification of certain proposals which appeared 
in the relation between the matrix models with massless flavors and gauge theory.

\subsection{Review of the results for pure gauge theories}

The large $N$ duality between open strings (branes) on the resolved geometry and 
closed strings (fluxes) on the deformed geometry was the 
starting point of the Dijkgraaf-Vafa conjecture. They argued that the effective superpotential of 
the gauge theory living on 
the non-compact part of D$5$ branes wrapping compact 2-cycles in the resolved geometry is given by 
the free energy of the matrix model built with the superpotential of the gauge theory and that this
free energy is equal to the one of the topological IIB superstrings on the deformed side.

In the case of the small resolution of the conifold, the arguments for this bold conjecture 
rely on the 
fact that the fields living on the 2-cycles are governed by the holomorphic Chern-Simons theory 
\cite{WIT92} as well as on the fact that a field theory superpotential
can be included in this theory by simply shifting the action by an amount equal to the product between 
the K\"ahler class and the superpotential evaluated on 0-form deformations \cite{KKLM}.
\be
Z=\int d\Phi_0d\Phi_1 e^{\int_{\cal C}\Phi_0{\bar D}\Phi_1+W(\Phi_0)\omega}
\ee
Then, the equations of motion allows one to set $\Phi_1=0$, as well as restrict 
$\Phi_0$ to the zero mode. Thus, the partition function reduces to just an integral over matrices:
\be
Z=\int d\Phi e^{-\frac{1}{g_s} {\Tr} W(\Phi)}
\label{matm1}
\ee
The assumptions of this proposal include the identification of the glueball superfield with the 't~Hooft 
coupling of the matrix model: $S = N g_s$.  It is important to emphasize that 
the dimension $N$ of the matrices appearing in the matrix model is unrelated to the rank of the gauge 
group $N_c$. 

The original Dijkgraaf-Vafa conjecture was stated for complex deformations of an 
${\cal N} = 2, A_1$  singularity.
Similar results hold in the case of other ${\cal N} = 2$ singularities, e.g. those which lead 
to quiver gauge 
theories. In that case, besides the integral of the adjoint fields as in (\ref{matm1}), some 
extra terms are required for integrating the bifundamental fields and the Chern-Simons action is 
simplified due to localization on the lowest lying modes.
Thus, one adds to the superpotential:
\be
W_{\it bif}=\sum_i\Tr[ Q_{i,i+1}\Phi_{i+1}{\tilde Q}_{i+1,i}]+ \Tr[{\tilde Q}_{i+1,i}\Phi_{i}{Q}_{i,i+1}]
\ee

We now turn to the discussion of fields in the fundamental representation and discuss the geometric 
justification of the various procedures of dealing with massless quarks \cite{janik, brr,feng}.

\subsection{Massive and massless matrix models} 

The geometry and matrix models for gauge theories with massive fundamental matter were related  
in several papers \cite{mcgrev,hofman,ookuochi} (see also \cite{dv2,seki} for the case of
bifundamental matter). In \cite{ookuochi} it was suggested that all the D5 branes are replaced by RR 
fluxes which would mean that all the 2-cycles shrink and are replaced by $S^3$ with flux. 

As described in section \ref{funnyS}, the simplest  way to deal, in a similar way, with both 
fundamental and bifundamental fields 
with  Yukawa-type superpotentials is to start 
with the product of  two gauge groups $U(N_c) \times U(N_f)$ with bifundamental fields 
and to take the coupling constant of the $U(N_f)$ group to zero. Thus, this symmetry becomes a global one
and the bifundamental fields transform in the fundamental or anti-fundamental 
representation of the remaining gauge group. This method was used in the DV context in 
\cite{hofman} and we will adjust it to our case. 

It is however worth pointing out that this limit can be interpreted 
geometrically as a ``partial geometric transition''. Indeed, the size of the $\P^1$ cycles wrapped by 
D$5$ branes
is proportional to the inverse of the coupling constant of the corresponding factor of the gauge group. Thus,
the geometric transition occurs only for the cycle wrapped by the branes generating the $U(N_c)$ gauge 
group,
while the others remain as $\P^1$ cycles; the vanishing coupling constant limit corresponds to 
decompactifying them.

Let us begin with the ${\cal N} = 2$ theory with product gauge group $U(N_c) \times U(N_f)$, 
which is geometrically engineered  as a resolved  $A_2$ singularity with $N_c$ D5 branes on one $\bP$ 
and $N_f$ D5 branes on the other $\bP$. We then break half of the supersymmetry to by 
adding a mass term for the adjoint fields. More generally, one can add an arbitrary potential for them,
but this does not modify the discussion. As discussed in section \ref{funnyS}, 
there is also an $S^3$ cycle  whose size is proportional to the vevs of the massless fundamental fields. 

The corresponding matrix model is \cite{dv2}:
\bea 
Z = \int~d\Phi_1~ d\Phi_2~dQ~d\tilde{Q} \,e^{- {\Tr} W(\Phi_1,\Phi_2,Q,\tilde{Q})}
\eea
where $\Phi_i$ are $M_i \times M_i$ matrices, $Q$ is $M_1 \times M_2$ matrix and
\bea
W(\Phi_1,\Phi_2,Q,\tilde{Q}) = {1\over g_1}W_1(\Phi_1) +  {1\over g_2}W_2(\Phi_2) + 
\\Tr[Q \Phi_2 \tilde{Q} - \tilde{Q} \Phi_1 Q]~~,
\eea
$W_i(\Phi_i)$ being polynomials of $\Phi_i$.

The above superpotential breaks supersymmetry to ${\cal N} = 1$ and the moduli space is described by the
expectations values for the adjoint fields $\Phi_i$ and for the bifundamental fields $Q$ and ${\tilde Q}$. 
Taking the limit of vanishing $g_2$ freezes $\Phi_2$ to one of the minima of $W_2$. Since we are 
interested in both massive and massless flavor fields, we assume that only part of the diagonal entries of 
$\Phi_2$ are nonvanishing. As the  diagonal entries of $\Phi_2$ give the mass for the quarks, we
have then a splitting of $Q$ and ${\tilde Q}$ into massive and massless fields, the potential
for the matrix model above being
\bea
V_{\it MM}(\Phi,Q,{\tilde Q})
=W(\Phi) + \sum_{i=1}^{N_f} (Q_i \Phi \tilde{Q}_i) + \sum_{i=1}^{N_j^m} m_{i} Q_i \tilde{Q}_i~~.
\eea

There is, however, more information which can be obtained from the description of massless flavors 
in the previous sections. In particular, we had to impose boundary conditions on 
the cycles describing both massive and massless flavors. The mass term in the previous equation
can be interpreted in that setup as arising because the noncompact 2-cycle describing massive 
fundamental fields ends at one of the nonvanishing minima of $W_2$. 

It is however clear that the above superpotential does not take into account the 
boundary conditions required 
for the cycle describing massless flavor fields. In the previous section we imposed the condition 
that the non-compact 2-cycles
describing the massless fields end on the curve $\Sigma$ which is orthogonal to 
them at the points fixed by the 
eigenvalues of the meson matrix. An equivalent way of stating this boundary condition is that the
holomorphic Chern-Simons action is evaluated with the constraint that the meson eigenvalues are fixed.

We can now supplement the potential $V_{\it MM}$ with the appropriate constraint. In the weak coupling
regime the meson is just a bilinear in quarks and antiquarks. 
Furthermore, one can perform an $SU(N_f)$ global rotation and replace the 
constraint that the eigenvalues of  $Q_i \tilde{Q}_j$ are fixed with the requirement that 
$Q_i \tilde{Q}_j$ is some fixed matrix. In principle, its eigenvalues should be equal to those 
which appear in the original constraint. However, as they were arbitrary, the corresponding 
matrix is arbitrary. Thus, the partition function of matrix model with massive and massless flavors is:
\be
Z={\cal N}\int d\Phi\, dQ^m d{\tilde Q}{}^m\,dQ {}^M d{\tilde Q}{}^M\delta(Q {}_i^M {\tilde Q}{}_j^M-M_{ij})
e^{-V_{\it MM}(\Phi,Q,{\tilde Q})}
\label{finmm}
\ee
The normalization of this partition function requires division by the inverse volume of the matrix model 
``gauge group'' $U(N)$. As in the original DV proposal, this can be interpreted as being
part of the planar and boundaryless free energy. Thus, its contribution to the superpotential 
is its derivative multiplied by the rank of the gauge group (the unbroken as well as the broken part!).

The equation (\ref{finmm}) recovers the suggestion \cite{janik} for the inclusion of massless quarks in 
the matrix model, and
is also equivalent \cite{feng} with the suggestion of \cite{brr} that one 
first deforms the matrix model by mass terms and then takes the massless limit. 
This can be easily seen by using an integral representation for the $\delta$-function  
in  equation (\ref{finmm}) and noticing that the new variable plays the role 
of mass parameter for the quarks $Q {}^M$ and ${\tilde Q}{}^M$.

For illustration purposes, let us briefly analyze the case of a quadratic superpotential for the 
adjoint field and
recover equation (\ref{Wfull}). We will also concentrate on the terms linear in $N_f$. In the cases 
covered by our analysis, this can be understood as arising from  the large $N$ limit for the matrix 
model gauge group.

The easiest way to go about computing the free energy in this regime is to represent it as a sum of vacuum
Feynman diagrams and furthermore notice that any diagram contains exactly one  species of quarks.
Thus, the free energy receives two independent contributions, one from the massive quarks and the other 
from the massless ones. The integral over  massive quarks is computed as in \cite{goteborg}
and gives
\be
Z=e^{-{1\over g_s} {\cal F}_{\it massive}}
\int d\Phi\, dQ {}^M d{\tilde Q}{}^M\delta(Q {}_i^M {\tilde Q}{}_j^M-M_{ij})
e^{-{1\over g_s}{{\Tr}}[Q^M\Phi{\tilde Q}^M + \frac{1}{2} m_A \Phi^2)]}
\ee
where
\bea
{\cal F}_{\it massive}&=&\sum_{i=1}^{N_f^m}{\cal F}(m_i)\\
{\cal F}(m)&=&S\left[-{1\over 2}-{1\over 4\alpha_m S}(\sqrt{1-4\alpha_m S}-1)
+\ln\left[{1\over 2}+{1\over 2}\sqrt{1-4\alpha_m S}\right]\right]\nonumber
\eea
and $\alpha_m={1\over m_Am^2}$.

The next step is to integrate out the adjoint field, as a gaussian integral which implies the 
appearance of $\Tr[(Q^M{\tilde Q}^M)^2]$. The remaining
integrand is then expressed only in terms of the bilinear $(Q^M{\tilde Q}^M)$ and it can 
be pulled out of the integral because of the $\delta$-function. Furthermore, this integral is 
also part of the planar and boundaryless free energy. As its contribution to the superpotential
is slightly different than the one of flavor fields, we will leave it aside for the moment. Surely enough,
we will add it back at the end. We are therefore left with:
\be
Z=e^{-{1\over g_s}\left({\cal F}_{\it massive}+{1\over m_Am^2}{\Tr}[M^2]\right)}
\int dQ {}^M d{\tilde Q}{}^M\delta(Q {}_i^M {\tilde Q}{}_j^M-M_{ij})
\ee
where $\Lambda$ is a cutoff introduced here for dimensional reasons. 

The remaining integral was performed in (\cite{janik})  and yields:
\be
\int dQ {}^M d{\tilde Q}{}^M\delta(Q {}_i^M {\tilde Q}{}_j^M-M_{ij})=
e^{{1\over g_s}\left[S\ln (\det M / \Lambda^{2N_f^M})-N_f^M S \ln {S\over \Lambda^3}\right]}~~.
\ee

Combining all the pieces together, the result is that the part of the free energy of 
the matrix model arising from the integration over fields is given by:
\be
-g_s\ln Z=\sum_{i=1}^{N_f^m}{\cal F}(m_i)+S\left[\ln{\Lambda^{2N_f^M}\over
\det M} -N_f^M\right]
\ee
where $\Lambda$ is a cutoff. Clearly the $U(N_f)$ invariance can be restored by replacing the 
product of eigenvalues of the meson matrix with its determinant.

Adding to this the contribution of the normalization coefficient ${\cal N}$ (i.e. the Veneziano-Yankielowicz
superpotential for the group $U(N_c)$) as well as the contribution of the integral over the adjoint field
recovers equation (\ref{fullWeff}), proving that the matrix model  (\ref{finmm}) describes the 
full nonperturbative physics of the gauge theory.

\section{Conclusions}

The main goal of our work was to fill certain gaps in understanding the relation between the 
gauge theories, geometry and matrix models for field theories with fields transforming in 
the fundamental representation of the gauge group. 

We described in detail the geometric construction of a supersymmetry-breaking mass term of finite size
for the adjoint field in the simplest ${\cal N}=2$ theory and analyzed the inclusion of massive and 
massless fields in the fundamental representation. The gauge theory described by this construction
is ${\cal N}=1$ SQCD with massive and massless quarks coupled with an adjoint field of finite mass 
through a Yukawa coupling. Analyzing the geometric transition for this construction we computed
the effective superpotential for this theory emphasizing the contribution of massless quarks as well as 
that of the finite adjoint mass.

Using the information we gained from this analysis we reconsidered the geometry prior to the
geometric transition. For theories without fields transforming in the fundamental representation
this was the starting point which lead to the DV proposal.  While the inclusion of massive quarks 
in this framework was easily achieved without reference to geometry, certain 
difficulties were encountered in dealing with massless ones. The two solutions to this problem, {\it 
proposed in \cite{janik} and \cite{brr} on a field-theoretic basis only}, were shown to be equivalent in 
\cite{feng}. From our analysis  we see that this identification appears naturally from the geometrical
picture and  its relation to brane configurations. As was emphasized before 
(see \cite{dot1}-\cite{dot4}), the brane configurations represent a very useful tool in the description 
of geometric transitions and even more so in the light of the new correspondences 
between geometry, field theories and matrix models. 

Finally, we illustrated the use of the matrix model we constructed and recovered the effective
superpotential computed from geometric and topological considerations and found
an exact agreement.

There are several directions which can be pursued further. As we describe the case of 
massless flavors, it would be interesting to use D6 branes instead of D4 branes. One immediate
problem is pushing them through the T-duality which gives a geometric description to the 
brane configuration, as naively they become D5 branes passing through the interior of the $P^1$, 
which does not belong to the space.

Another interesting direction is to consider brane configurations corresponding to chiral theories;
T-duality transformations would map them to geometries which cannot be obtained from ${\cal N} = 2$ 
ones by deformations. If possible, these would become an extension of the original conjectures to the
chiral case. For many models of brane configurations see \cite{GIKU}. 

\vspace*{2cm}

\noindent
{\bf Acknowledgments}

\noindent
R.T. would like to thank Keshav Dasgupta and Kyungho Oh for discussions 
and collaboration on related issues. R.R. and J.W. thank Oliver DeWolfe for
discussions. R.R.\ is supported in part by DOE under 
Grant No.\ 91ER40618 and in part by the NSF under Grant No.\ PHY00-98395.
R.T.\ is supported by the DOE grant No.\ DE-AC03-76SF00098, the NSF under Grant No.\ PHY-0098840 
and by the Berkeley Center for Theoretical Physics. 
J.W.\ is supported in part by the NSF under Grant No.\ PHY99-07949.
Any opinions, findings, and conclusions or recommendations expressed in this material
are those of the authors and do not necessarily reflect the views
of the National Science Foundation.


\begin{thebibliography}{9}

\bibitem{gv} 
R.~Gopakumar and C.~Vafa,
Adv.\ Theor.\ Math.\ Phys.\  {\bf 3}, 1415 (1999),
hep-th/9811131.

\bibitem{vafa}
C.~Vafa,
J.\ Math.\ Phys.\  {\bf 42}, 2798 (2001), hep-th/0008142.

\bibitem{CIVA} 
F.~Cachazo, K.~A.~Intriligator and C.~Vafa,
Nucl.\ Phys.\ B {\bf 603}, 3 (2001), hep-th/0103067.

\bibitem{eot}
J.~D.~Edelstein, K.~Oh and R.~Tatar,
JHEP {\bf 0105}, 009 (2001), hep-th/0104037.

\bibitem{ckv}
F.~Cachazo, S.~Katz and C.~Vafa,
hep-th/0108120.

\bibitem{civ1}
F.~Cachazo, B.~Fiol, K.~A.~Intriligator, S.~Katz and C.~Vafa,
Nucl.\ Phys.\ B {\bf 628}, 3 (2002), hep-th/0110028.

\bibitem{fiol}
B.~Fiol,
JHEP {\bf 0207}, 058 (2002), hep-th/0205155.

\bibitem{ook1} 
H.~Fuji and Y.~Ookouchi,
hep-th/0205301.

\bibitem{cv} 
F.~Cachazo and C.~Vafa,
hep-th/0206017.

\bibitem{dot1}
K.~Dasgupta, K.~Oh and R.~Tatar,
Nucl.\ Phys.\ B {\bf 610}, 331 (2001), \\
hep-th/0105066.

\bibitem{dot2}
K.~Dasgupta, K.~Oh and R.~Tatar,
JHEP {\bf 0208}, 026 (2002), \\
hep-th/0106040.

\bibitem{dot3}
K.~h.~Oh and R.~Tatar,
hep-th/0112040, Adv. Theor. Math. Phys. {\bf 6} (2002) 141-196

\bibitem{dot4}
K.~Dasgupta, K.~h.~Oh, J.~Park and R.~Tatar,
JHEP {\bf 0201}, 031 (2002), hep-th/0110050.

\bibitem{dv1} 
R.~Dijkgraaf and C.~Vafa,
Nucl.\ Phys.\ B {\bf 644}, 3 (2002), hep-th/0206255.

\bibitem{dv2}
R.~Dijkgraaf and C.~Vafa,
Nucl.\ Phys.\ B {\bf 644}, 21 (2002), hep-th/0207106.

\bibitem{dv3} 
R.~Dijkgraaf and C.~Vafa,
hep-th/0208048.

\bibitem{grisaru}
R.~Dijkgraaf, M.~T.~Grisaru, C.~S.~Lam, C.~Vafa and D.~Zanon,
hep-th/0211017.

\bibitem{witten}
F.~Cachazo, M.~R.~Douglas, N.~Seiberg and E.~Witten,
arXiv:hep-th/0211170.

\bibitem{csw}
F.~Cachazo, N.~Seiberg and E.~Witten,
arXiv:hep-th/0301006.

\bibitem{Berenstein}
D.~Berenstein,
Phys.\ Lett.\ B {\bf 552}, 255 (2003),
hep-th/0210183.

\bibitem{goteborg}
R.~Argurio, V.~L.~Campos, G.~Ferretti and R.~Heise,
hep-th/0210291.

\bibitem{mcgrev} 
J.~McGreevy,
hep-th/0211009.

\bibitem{Suzuki}
H.~Suzuki,
hep-th/0211052,
hep-th/0212121.

\bibitem{janik} 
Y.~Demasure and R.~A.~Janik,
hep-th/0211082.

\bibitem{rr} 
I.~Bena and R.~Roiban,
hep-th/0211075, to appear in Phys.Let.{\bf B}.

\bibitem{Tachikawa}
Y.~Tachikawa,
hep-th/0211189,
hep-th/0211274.

\bibitem{Feng}
B.~Feng, 
hep-th/0211202;
B.~Feng and Y.~H.~He,
hep-th/0211234

\bibitem{Argurio}
R.~Argurio, V.~L.~Campos, G.~Ferretti and R.~Heise,
Phys.\ Lett.\ B {\bf 553}, 332 (2003), hep-th/0211249.

\bibitem{Naculich:2002hr}
S.~G.~Naculich, H.~J.~Schnitzer and N.~Wyllard,
JHEP {\bf 0301}, 015 (2003), hep-th/0211254; 
hep-th/0211123.

\bibitem{brr}
I.~Bena, R.~Roiban and R.~Tatar,
hep-th/0211271.

\bibitem{ookuochi}
Y.~Ookouchi,
hep-th/0211287.

\bibitem{Ohta:2002rd}
K.~Ohta,
hep-th/0212025.

\bibitem{seki}
S.~Seki,
hep-th/0212079.

\bibitem{Bena:2002tn}
I.~Bena, S.~de Haro and R.~Roiban,
hep-th/0212083.

\bibitem{hofman}
C.~Hofman,
hep-th/0212095.

\bibitem{Demasure:2002jb}
Y.~Demasure and R.~A.~Janik,
hep-th/0212212.

\bibitem{seib} 
N.~Seiberg,
hep-th/0212225.

\bibitem{Ahn:2002vj}
C.~Ahn and S.~Nam,
hep-th/0212231; C. Ahn, 
hep-th/0301011.

\bibitem{feng} 
B.~Feng,
arXiv:hep-th/0212274.

\bibitem{mt}
F.~Ferrari,
Nucl.\ Phys.\ B {\bf 648}, 161 (2003), hep-th/0210135;\\
hep-th/0211069.

H.~Fuji and Y.~Ookouchi,
JHEP {\bf 0212}, 067 (2002), hep-th/0210148:

R.~Dijkgraaf, S.~Gukov, V.~A.~Kazakov and C.~Vafa,
hep-th/0210238.

A.~Gorsky,
hep-th/0210281.

R.~Gopakumar,
hep-th/0211100.

R.~Dijkgraaf, A.~Neitzke and C.~Vafa,
hep-th/0211194.

A.~Klemm, M.~Marino and S.~Theisen,
hep-th/0211216.

R.~Dijkgraaf, A.~Sinkovics and M.~Temurhan,
hep-th/0211241.

H.~Itoyama and A.~Morozov,
hep-th/0211245.

S.~K.~Ashok, R.~Corrado, N.~Halmagyi, K.~D.~Kennaway and C.~Romelsberger,
hep-th/0211291.

B.~Feng,
hep-th/0212010.

H.~Itoyama and A.~Morozov,
hep-th/0212032, 
hep-th/0301136.

R.~A.~Janik and N.~A.~Obers,
Phys.\ Lett.\ B {\bf 553}, 309 (2003)
hep-th/0212069.

V.~Balasubramanian, J.~de Boer, B.~Feng, Y.~H.~He, M.~x.~Huang, V.~Jejjala and A.~Naqvi,
hep-th/0212082.

M.~Matone,
hep-th/0212253.

A.~Dymarsky and V.~Pestun,
hep-th/0301135.

F.~Ferrari,
hep-th/0301157.



\bibitem{seiberg}
N.~Seiberg,
Phys.\ Rev.\ D {\bf 49}, 6857 (1994), hep-th/9402044.

\bibitem{INTR} 
K.~A.~Intriligator,
Phys.\ Lett.\ B {\bf 336}, 409 (1994), hep-th/9407106].

\bibitem{AGVA} 
M.~Aganagic and C.~Vafa,
arXiv:hep-th/0012041.

\bibitem{hw}
A.~Hanany and E.~Witten,
Nucl.\ Phys.\ B {\bf 492}, 152 (1997),\\
 hep-th/9611230.

\bibitem{wit} 
E.~Witten,
Nucl.\ Phys.\ B {\bf 500}, 3 (1997), hep-th/9703166.


\bibitem{OOHO} 
K.~Hori, H.~Ooguri and Y.~Oz,
Adv.\ Theor.\ Math.\ Phys.\  {\bf 1}, 1 (1998), hep-th/9706082.

\bibitem{BARB}
J.~L.~Barbon,
Phys.\ Lett.\ B {\bf 402}, 59 (1997); hep-th/9703051.

\bibitem{itzaki} 
A.~Brandhuber, N.~Itzhaki, V.~Kaplunovsky, J.~Sonnenschein and S.~Yankielowicz,
Phys.\ Lett.\ B {\bf 410}, 27 (1997), hep-th/9706127.

\bibitem{GIKU}
A.~Giveon and D.~Kutasov,
Rev.\ Mod.\ Phys.\ {\bf 71}, 983 (1999), \\
hep-th/9802067.


\bibitem{oova} H.~Ooguri and C.~Vafa,
Nucl.\ Phys.\ B {\bf 463}, 55 (1996), hep-th/9511164.

\bibitem{bh} 
J.~H.~Brodie and A.~Hanany,
Nucl.\ Phys.\ B {\bf 506}, 157 (1997), \\
hep-th/9704043.

\bibitem{WIT92} 
E.~Witten,
Prog.\ Math.\  {\bf 133}, 637 (1995), hep-th/9207094.


\bibitem{DWFO} O.~DeWolfe, D.Z.~Freedman and H.~Ooguri,  
Phys.Rev.D66:025009,2002; hep-th/0111135 

\bibitem{gns}
S.~Gubser, N.~Nekrasov and S.~Shatashvili,
JHEP {\bf 9905}, 003 (1999), hep-th/9811230.

\bibitem{KKLM} S.~Kachru, S.~Katz, A.~Lawrence and J.~McGreevy\\
Phys.\ Rev.\ D {\bf 62}, 026001 (2000), hep-th/9912151.

\bibitem{GIPE} A.~Giveon and O.~Pelc 
Nucl.\ Phys.\ B {\bf 512}, 103 (1998), hep-th/9708168.





\end{thebibliography}
\end{document}